# Effect of radiation from vacuum in the interaction of zero-point fluctuations with charged particles accelerated in electric or magnetic fields

Ya. B. Ulanovsky, A.M. Frolov

**Abstract**

The possibility of the effect of electromagnetic radiation (hereinafter - radiation from vacuum) due to the interaction of zero-point vacuum fluctuations with free charged particles accelerated at relativistic speeds in electric or magnetic fields is theoretically demonstrated and confirmed by calculations. The proposed effect is similar to the appearance of radiation in the Lamb shift effect in the hydrogen atom and the dynamical Casimir effect. However, unlike these effects, the radiation characteristics depend on the induction of electric or magnetic fields. Moreover, in order to comply with the law of conservation of momentum, the braking radiation of charged particles is considered as the "third body". As opposed to this bremsstrahlung, radiation from vacuum will not have a preferred direction.

Analytical expressions describing the radiation characteristics were obtained using a semi-classical method. The calculated values of the energy and intensity of radiation from vacuum demonstrate the possibility of using optical methods to determine the characteristics of such radiation in strong electric or magnetic fields. The results of the work can be used in studies of astrophysical objects with powerful electric or magnetic fields, as well as in investigations of physical vacuum states under laboratory conditions.

Keywords: **zero-point oscillations, electrons, photons, bremsstrahlung, vacuum, magnetic field, electric field, Lamb shift effect, effect of radiation, physical effect**

**Introduction**

The possible effect of electromagnetic radiation (hereinafter - radiation from vacuum) due to the interaction of zero-point vacuum fluctuations with free charged particles accelerated at relativistic speeds in electric or magnetic fields is considered.

It is known that the interaction[a] of zero-point vacuum fluctuations with charged particles can lead to the transformation of such zero-point vacuum fluctuations into real photons. This phenomenon can be observed in the Lamb shift effect in the hydrogen atom [1,2] and the dynamical Casimir effect [3,4]. As the zero-point vacuum fluctuations have no momentum, only the interaction with a charged particle is insufficient for their conversion into real photons. In this case, in order to comply with the law of conservation of momentum [1–3], the charged particle must transmit momentum not only to zero-point oscillations, but also to a so-called "third body[b]".

According to the Lamb shift effect in a hydrogen atom, such a "third body" to which momentum can be transmitted consists of the electric field of the nucleus – or, otherwise stated, the nucleus itself. In terms of the dynamical Casimir effect, such a "third body" is envisaged as a lattice of metal ions. We will apply an analogous approach to consider the possible transformation of zero-point vacuum fluctuations into real photons when the former come under the action of free charged particles accelerating relativistically in electric or magnetic fields. Assuming the possibility of such a transformation, the authors of the present work consider that, in this case, the "third body" can be the bremsstrahlung (braking radiation) of charged particles if their momentum has a nonzero value. This occurs if an accelerated electron has a relativistic velocity. In this case, due to the relativistic effect, the expression for the bremsstrahlung contains nonlinear terms: the spatial distribution of the braking radiation becomes heterogeneous and the total momentum of the bremsstrahlung ceases to be equal to 0.

Therefore, by interacting with zero-point vacuum fluctuations, the electron can transmit momentum to the bremsstrahlung and hence to the zero-point vacuum fluctuations.

At the same time, unlike bremsstrahlung, the radiation from vacuum will not have a preferred direction since emission of photons from vacuum is equally probable in all directions relative to the trajectory of an electron. This makes it possible to record such radiation and its characteristics – i.e., the intensity of radiation and its spectrum – as well as to compare it with the calculated data presented in this article using sensors aimed at an electron's trajectory from any side, including following the direction of the electron's movement.

The distinguishing feature of the proposed effect in comparison with the similar Lamb shift and dynamical Casimir effects is the consideration of the bremsstrahlung of charged particles as the "third body".

**Calculations of parameters of radiation from vacuum.**

Below are calculations of the parameters of radiation from vacuum during the interaction of zero-point vacuum fluctuations with free electrons accelerating at relativistic speeds in electric or magnetic fields. First, a nonlinear component is extracted from the general expression for the bremsstrahlung, followed by averaging by zero-point vacuum fluctuations and converting the resulting expression into an algebraic expression. By analogy, the same calculations can be performed for other charged particles. Calculations are carried out in the SI system using the semi-classical method.

The equation of motion of an electron in an electric field has the form [5]:
$$m_e(d\mathbf{V}/dt) = e\mathbf{E}/\varepsilon_0, \quad (1)$$
where $\varepsilon_0$ – vacuum permittivity; $m_e$ – electron mass; e – electron charge; $\mathbf{E}$ – electric field strength vector; $\mathbf{V}_E = d\mathbf{r}_E/dt$ – velocity vector of an electron in an electric field.

The motion of an electron in a magnetic field has the form [3]:
$$m_e(d\mathbf{V}_H/dt) = (e/c)[\mathbf{HV}], \quad (2)$$
where $\mathbf{H}$ – magnetic field intensity vector; $\mathbf{V}_H = d\mathbf{r}_H/dt$ – velocity vector of the electron in a magnetic field.

With accelerated electron motion, the bremsstrahlung is generally understood as determined by the equation [6]:
$$-d\varepsilon/dt = (2/3)(e^4/(\varepsilon_0 m_e^2 c^3))((\mathbf{E} + [\mathbf{VH}]/c)^2 - (\mathbf{EH})/c^2)/(1-(V/c)^2)^2 \quad (3)$$

As a result of substituting equations (1) and (2) into equation (3), we obtain:
$$-d\varepsilon/dt = (2/3)(e^2/c^3)(d\mathbf{V}/dt)^2 - [\mathbf{V}(d\mathbf{V}/dt)]^2/c^2)/(1-(V/c)^2)^3 \quad (3')$$

In this case, the change in the bremsstrahlung momentum is not equal to zero, since it is associated with equation (3) by the ratio $-d\mathbf{P}/dt = -(\mathbf{V}/c^2)d\varepsilon/dt$. Consequently, the bremsstrahlung can transmit momentum through the electron to zero-point electromagnetic vacuum fluctuations. Equation (3) is applicable in our case given that, in the rest frame, the vector momentum of the electron is not zero.

The zero-point vacuum fluctuations themselves, comprising electromagnetic oscillations, interact with the electron, causing it to fluctuate spatially throughout the entire trajectory when moving in vacuum. Accordingly, these spatial displacements of the electron trajectory under the influence of zero-point fluctuations should be determined by the same equation as zero-point vacuum fluctuations [7] in order to allow the exchange of energy and momentum between the two types of oscillations:

$$\overline{\delta r^2} = (2/\pi)\alpha(\hbar/m_e c)^2 \ln(\omega_c/(m_e c^2/\hbar)), \qquad (4)$$

where $\alpha = e^2/\hbar c$ – fine structure constant; $\hbar$ – Planck constant.

When an electron is simultaneously exposed to zero-point electromagnetic fluctuations of vacuum and braking radiation, the trajectory of the electron at each point will change by the magnitude of the fluctuation $\delta r$. This spatial fluctuation leads to the appearance of electron dipole radiation and, accordingly, to the simultaneous occurrence of fluctuations in the bremsstrahlung as a result of the appearance of radiation from the vacuum. Proceeding from equations (3) and (4), we determine the relationship between the spatial fluctuation $\delta r$ and the energy fluctuation $\delta\varepsilon$ separately for electric and magnetic fields. In turn, zero-point electromagnetic vacuum fluctuations lead to fluctuations in the intensity of the bremsstrahlung depending on the spatial displacement fluctuations of the accelerated electron. In order to determine such a dependence of the bremsstrahlung fluctuation for the electron in the electric field, we proceed from the fact that the bremsstrahlung energy and electron velocity fluctuations are infinitesimal compared to the bremsstrahlung and electron velocity magnitudes, replacing $d\varepsilon/dt$ by $d\varepsilon/dt + d'v_e\varepsilon/dt$ and $\mathbf{V}$ by $\mathbf{V}+ $ and$\mathbf{V}$ in equation (3) and substituting from equation (1) the value for the acceleration $d\mathbf{V_E}/dt$ and the average for all zero-point vacuum oscillations. As a result, we obtain an equation consisting of the sum of the bremsstrahlung energy and the radiant vacuum energy:

$$- d\varepsilon_E/dt + d\delta\overline{\varepsilon}_E/dt, \qquad (5)$$

where the first term corresponds to the bremsstrahlung of an electron in an electric field (3),

$$- d\varepsilon_E/dt = (2/3)(e^4/(\varepsilon_0 m_e^2 c^3))((\mathbf{E})^2/(1 - (V/c)^2)^2$$

and the second term equals

$$d\delta\overline{\varepsilon}_E/dt = \sin\beta(2/3)(e^4/\varepsilon_0 m_e^2 c^3)E^2 d\delta\overline{V}^2_E/dt/c^2 \qquad (6)$$

corresponding to the radiation from vacuum.

Performing analogous transformations for the bremsstrahlung of an electron in a magnetic field using equation (2), we obtain:

$$- d\varepsilon_H/dt + d\delta\overline{\varepsilon}_H/dt, \qquad (7)$$

where the first term corresponds to the bremsstrahlung of an electron in a magnetic field (3),

$$- d\varepsilon/dt = (2/3)(e^4/(\varepsilon_0 m_e^2 c^3))([\mathbf{VH}]/c)^2)/(1 - (V/c)^2)^2$$

while the second term
$$d\delta\bar{\varepsilon}_H/dt = \sin\beta (2/3)(e^4/\varepsilon_0 m_e^2 c^3)H^2 d\delta\bar{V}^2_H/dt/c^2 \qquad (8)$$
corresponds to the radiation from vacuum.
(β is the angle between the electron velocity vector and the field lines of force)
Here equations (6) and (8) no longer define braking processes, but processes of radiation from vacuum, since these values, unlike the values of bremsstrahlung, have positive values.

Now we will transform equations (6) and (8) into algebraic equations. To do this, we substitute the time derivatives in the left and right parts of these equations with the electron radiation frequencies separately for each field $\omega_E$ and $\omega_H$. For the electric field, we obtain an equation in which the relationship of the fluctuation $\delta\varepsilon_E$ and $\delta r^2_E$ is determined:
$$\delta\bar{\varepsilon}_E = (2/3)(e^4/\varepsilon_0 m_e^2 c^4)E^2 \delta\bar{r}^2_E (\omega/c), \qquad (9)$$
while the relationship of fluctuations $\delta\varepsilon_H$ and $\delta r^2_H$ for the magnetic is defined field similarly:
$$\delta\bar{\varepsilon}_H = (2/3)(e^4/\varepsilon_0 m_e^2 c^4)H^2 \delta\bar{r}^2_H (\omega/c) \qquad (10)$$
Now we transform equations (9) and (10), in which the variables $\varepsilon_E$, $\varepsilon_H$ and $r_E$, $r_P$ are interconnected, but already in the form of fluctuations, into algebraic equations. To do this, we first replace the time derivatives with the bremsstrahlung frequencies for electron radiation in the left and right parts of these equations (9) and (10) separately for each field $\omega_E$ and $\omega_H$. Next, we perform frequency averaging to obtain an equation in which the relationship of the fluctuations $\delta\varepsilon_E$ and $\delta r_E$ is determined for the electric field. From equation (9), we obtain a relation that connects two unknown variables for the electric field - E and $\omega_E$:
$$\omega_E = \delta\varepsilon_E/\hbar = (2/3)(c/\omega_E \hbar)(e^4/(\varepsilon_0 m_e^2 c^4))E^2, \qquad (11)$$
from where we get the functional dependence between E and $\omega_E$ in the form:
$$\omega_E = ((2/3)(c/\hbar)(e^4/\varepsilon_0 m_e^2 c^4))^{1/2} E = 10^{12} E \qquad (12)$$
From equation (10), we obtain the relation that connects two unknown variables for the magnetic field - H and $\omega_H$:
$$\omega_H = \delta\varepsilon_H/\hbar = (2/3)(c/\omega_H \hbar)(e^4/(\varepsilon_0 m_e^2 c^4))H^2, \qquad (13)$$
from which we get the functional dependence between H and $\omega_H$ in the form:
$$\omega_H = ((2/3)(c/\hbar)(e^4/\varepsilon_0 m_e^2 c^4))^{1/2} H = 10^{12} H \qquad (14)$$
In order to isolate the energy fluctuation only for radiation from vacuum, we will average all zero-point vacuum fluctuations in equations (9) and (10). Hence, we obtain a new equation for the energy fluctuation of bremsstrahlung in the electric field $\delta\varepsilon_E$ depending on the spatial fluctuations $\delta r^2_E$ and bremsstrahlung $\omega_E$ in the electric field:
$$\delta\bar{\varepsilon}_E = (2/3)(e^4/\varepsilon_0 m_e^2 c^4)E^2 \delta\bar{r}^2_E (\omega_E/c) \qquad (15)$$
and, analogically, the equation for the energy fluctuation of braking radiation in the magnetic field $\delta\bar{\varepsilon}_H$ depending on the spatial fluctuations $\delta\bar{r}^2_H$ and bremsstrahlung $\omega_H$ in the magnetic field:
$$\delta\bar{\varepsilon}_H = (2/3)(e^4/\varepsilon_0 m_e^2 c^4)H^2 \delta\bar{r}^2_H (\omega_H/c) \qquad (16)$$
Let us determine the values of radiant energy from vacuum for an electron accelerated with relativistic velocity in an electric or magnetic field. Considering that an electron simultaneously interacts not only with an electric or magnetic field, but also with a vacuum through zero-point vacuum fluctuations, it is necessary to take into account the effect caused by the permittivity of the vacuum under the action of an electromagnetic field, determined by ratios $D = \varepsilon_0 E$ and $B = \mu_0 H$. Using these relations and substituting in (15) and (16) respectively $\delta r^2{}_E$ and $\delta r^2{}_H$ from (4), as well as $\omega_E$ and $\omega_H$ from (12) and (14), we obtain equations for determining the radiant energy from vacuum in electric and magnetic fields:

$$\delta\bar{\varepsilon}_E = ((2/3)(e^4/(\varepsilon_0 m_e^2 c^4))^{3/2} (D/\varepsilon_0)^3 (1/c\hbar)^{1/2}(2/\pi)\alpha(\hbar/m_e c)^2 \ln(v_e/\omega_E) \quad (17)$$

$$\delta\bar{\varepsilon}_H = ((2/3)(e^4/(\varepsilon_0 m_e^2 c^4))^{3/2} (B/\mu_0)^3 (1/c\hbar)^{1/2}(2/\pi)\alpha(\hbar/m^e c)^2 \ln(v_e/\omega_H), \quad (18)$$

where $\varepsilon_0$ – dielectric permittivity of vacuum; $\mu_0$ – magnetic vacuum permeability; $v_e = \hbar/m_e c^2$ – electronic vacuum frequency.

Substituting the known values of physical constants in the SI system into equations (17) and (18), we obtain the dependencies of energy and intensity for radiation from vacuum in electric and magnetic fields:

$$\delta\bar{\varepsilon}_{Evac} = 10^{-34} D^3 \text{ и } d\bar{\varepsilon}_{Evac}/dt = \delta\bar{\varepsilon}_{Evac}\omega_E = 10^{-22} D^4, \quad (19)$$

$$\delta\bar{\varepsilon}_{Hvac} = 10^{-25} B^3 \text{ и } d\bar{\varepsilon}_{Hvac}/dt = \delta\bar{\varepsilon}_{Hvac}\omega_H = 10^{-7} B \quad (20)$$

From equations (19) and (20), it follows that the energy and intensity of radiation from vacuum depend on the magnitude of the induction of the corresponding field. If the value of the electric field induction $D = 3 \cdot 10^4$ C/m$^2$, then the values of energy and radiation intensity from the vacuum are equal to $\sim 3 \cdot 10^{-21}$ J and $\sim 10^{-4}$ J/s, respectively. If the value of magnetic field induction $B = 10^2$ tf, then the energy and intensity of radiation from the vacuum, as follows from (20), are equal to $\sim 10^{-19}$ J and $\sim 10$ J/s, respectively, which corresponds to a frequency of radiation from the vacuum of $10^5$ GHz. These values are much higher than the energy and intensity of electron radiation at the Lamb shift, which are $\sim 10^{-24}$ J and $10^{-16}$ J/s. In the case of the Lamb shift effect, the radiation frequency is 1 GHz.

**Conclusion**

Along with the obtained analytical expressions (17), (18), the calculations (19), (20) performed in this work confirm the possibility of an electromagnetic radiation – radiation from vacuum effect. Such radiation occurs when both bremsstrahlung and zero-point vacuum fluctuations are simultaneously affected by an electron accelerated with relativistic velocity.

In should be noted that when the charged particles pass near astrophysical objects having a strong magnetic field, the effect of radiation from vacuum can be observed also in the optical range. The obtained results on the effect of converting zero-point vacuum fluctuations into real photons under the action of bremsstrahlung momentum of charged particles accelerated at relativistic speeds in an electric or magnetic field serve as an additional proof of the existence of phenomena such as the effect of zero-point vacuum fluctuations on charged particles in accordance with the laws of quantum electrodynamics (QED).

**References**

1. Berestetsky V.B., Akhiezer A.I., Quantum electrodynamics, Moscow, Nauka, 2002, p. 335
2. 3. F. Barranco[1], G. Potel[2], E. Vigezzi[3], and R. A. Broglia, arXiv:1904.02786v1 [nucl-th] 4 Apr 2019

3. V. V. Dodonov Current status of the Dynamical Casimir Effect, arXiv:1004.3301v2 [quant-ph] *[Submitted on 19 Apr 2010 (v1), last revised 24 Oct 2010 (this version, v2)]*
4. G.V. Dedkov, A.A. Kyasov The dynamic Casimir–Polder force in relativistic motion of an atom near the surface of a thick plate, Solid State Physics, 2012, volume 54, issue 4, p.781
5. Toptygin I.N. Modern electrodynamics Part 1. 2002, pp. 311- 437.
6. L.D. Landau, E.M. Lifshits, Field Theory, vol.2, 1988, pp. 247–285
7. A. B. Migdal "Qualitative methods in quantum theory", M., Nauka, 1975, ch. 1, pp. 68–71